\documentclass[aps,twocolumn,showpacs,showkeys,amsmath,amssymb,prb]{revtex4} 

\usepackage{graphicx}   
\usepackage{dcolumn}    
\usepackage{bm}         
\usepackage{natbib}

\newcommand{\rd}{{\rm d}} 
\newcommand{\re}{{\rm e}} 
\newcommand{\ri}{{\rm i}} 
\newcommand{\R}{\mathbb{R}}
\newcommand{\bnab}{{\bf \nabla}}
\newcommand{\bfn}{{\bf n}}
\newcommand{\bfN}{{\bf N}}
\newcommand{\bfndot}{{\bf\dot{n}}}
\newcommand{\bfr}{{\bf r}}
\newcommand{\udot}{\dot{u}}
\newcommand{\nablad}{\bnab_\partial}
\newcommand{\Deltad}{\Delta_\partial}
\renewcommand{\div}{\operatorname{div}}
\newcommand{\divd}{\div_\partial}
\newcommand{\bk}{\boldsymbol{\kappa}}

\begin{document} 
  
\title{Perturbation theory for plasmonic eigenvalues}

\author{Daniel Grieser}  
\author{Hannes Uecker}
\affiliation{Institut f\"ur Mathematik, Carl von Ossietzky Universit\"at, 
	D-26111 Oldenburg, Germany}
\author{Svend-Age Biehs\footnote{Present address: Laboratoire Charles Fabry,  
	Institut d'Optique, CNRS, Universit\'{e} Paris-Sud, 
	Campus Polytechnique, RD128, 91127 Palaiseau cedex, France}}
\author{Oliver Huth}
\author{Felix R\"uting}
\author{Martin Holthaus}
\affiliation{Institut f\"ur Physik, Carl von Ossietzky Universit\"at, 
	D-26111 Oldenburg, Germany}
\date{October 27, 2009}

\begin{abstract}
We develop a perturbative approach for calculating, within the quasistatic 
approximation, the shift of surface resonances in response to a deformation 
of a dielectric volume. Our strategy is based on the conversion of the 
homogeneous system for the potential which determines the plasmonic
eigenvalues into an inhomogeneous system for the potential's derivative
with respect to the deformation strength, and on the exploitation of the 
corresponding compatibility condition. The resulting general expression 
for the first-order shift is verified for two explicitly solvable cases,
and for a realistic example of a deformed nanosphere. It can be used for
scanning the huge parameter space of possible shape fluctuations with only 
quite small computational effort.    
\end{abstract}




\maketitle
 
\section{Introduction}

Recent advances in plasmonics, {\em i.e.\/}, in the use of surface plasmons
for subwavelength optics,~\cite{BarnesEtAl03,MaierAtwater05,Ozbay06} have
led to renewed interest in the physics of plasmon excitations bound to 
smooth or rough surfaces.~\cite{Raether88} In particular, it has been 
proposed to transport electromagnetic energy along linear chains of 
nanoparticles,~\cite{QuintenEtAl98} possibly embedded in a gain 
medium,~\cite{Citrin06} and proof-of-principle experiments have been
made.~\cite{BrongersmaEtAl00, MaierEtAl01,MaierEtAl03} In such setups, 
fabrication-induced shape imperfections of the nanoparticles inevitably  
will result in slight shifts of their resonance frequencies, and it might 
be crucial to estimate the maximum size of such fluctuations which can 
still be tolerated.~\cite{deWaeleEtAl07} Although sophisticated numerical 
methods for computing the optical response of nanoparticles do 
exist,~\cite{MartinEtAl95,HohenesterKrenn05} it would still be helpful to 
have a flexible analytical tool which exploits the fact that the shape of 
an unintentionally deformed nanoparticle is close to some theoretical ideal, 
as this tool would allow one to explore the huge parameter space of possible
perturbations in a computationally cheap manner.   

In the present paper we develop such a perturbative approach to the 
computation of surface resonances. Our strategy is quite general, relying 
on concepts borrowed from differential geometry.~\cite{doCarmo76} The  
mathematical arguments are given in the following Sec.~\ref{S_pert}; some 
technical details have been deferred to Appendix~\ref{App1}. The main results 
are the expressions~(\ref{RES}) and (\ref{VAR}), which quantify the shifts 
of the resonance frequencies to first order in the deformation strength. 
Section~\ref{S_exam} then provides three illustrative examples. The first 
two of these make contact with analytically solvable models, thus helping to 
gain confidence in the formalism, while the third one is a more realistic 
application to a deformed nanosphere for which no closed analytical solution 
is available. This necessitates to consider the splitting of degenerate modes,
which is done in Appendix~\ref{App2}. The paper ends with some concluding 
remarks in Sec.~\ref{S_conc}.

\section{Perturbation theory for plasmonics}
\label{S_pert}
 
Consider a volume $\Omega \subset \R^3$ filled with a dielectric medium 
which is characterized by an isotropic, frequency-dependent dielectric 
function $\epsilon(\omega)$; this volume $\Omega$ be surrounded by vacuum.
We employ the quasistatic approximation, which is valid if the characteristic 
linear extensions of $\Omega$ are small in comparison with the wavelengths 
of impinging radiation, and thus captures essential features of nonretarded 
plasmon dynamics in small nanoparticles.~\cite{HohenesterKrenn05,LiEtAl03,
ParkStroud04,NordlanderEtAl04,KlimovGuzatov07} The Fourier components 
$u(\bfr | \omega)$ of the potential then are given by the solutions to the 
set of equations
\begin{eqnarray}
	\Delta u & = & 0  \quad\text{in }\R^3\setminus\partial\Omega
\label{EQ1}
\\
	u_- - u_+ & = & 0 \quad\text{on }\partial\Omega
\label{EQ2}
\\
	\epsilon \partial_n u_- - \partial_n u_+ & = & 0 
	\quad\text{on } \partial\Omega \; .
\label{EQ3}	
\end{eqnarray}
Here $u_-$ is the restriction of $u$ to the interior $\Omega$ and $u_+$
its restriction to the open exterior $\R^3\setminus \overline\Omega$; both 
$u_-$ and $u_+$ are smooth everywhere except at the boundary $\partial\Omega$. 
We stipulate that $\partial\Omega$ be sufficiently smooth to substantiate 
the following operations, and suppress the dependence of $u$ on the 
frequency $\omega$ altogether. The expressions 
$\partial_n u_\pm = \bfn \cdot \bnab u_\pm$ in Eq.~(\ref{EQ3}) 
are the derivatives of the potential in the direction of the outward 
unit normal $\bfn$. 

Observe that at this point the knowledge of the dielectric function 
$\epsilon(\omega)$ is not yet required. Rather, we take $\epsilon$
on the l.h.s.\ of Eq.~(\ref{EQ3}) as a real number and regard it as
an eigenvalue, henceforth dubbed as {\em plasmonic eigenvalue\/}, if
the system~(\ref{EQ1}) -- (\ref{EQ3}) with that particular $\epsilon$ 
possesses a nontrivial solution~$u$ which vanishes at least quadratically 
with increasing distance from $\partial \Omega$. This nomenclature is
formally justified by reformulating the system~(\ref{EQ1}) -- (\ref{EQ3}) 
in terms of Dirichlet-to-Neumann operators, so that the desired values of 
$\epsilon$ explicitly appear as inverse eigenvalues of a combination of
such operators.~\cite{GrieserRueting09}
This somewhat unfamiliar view offers conceptual advantages:
On the one hand, the material-specific aspects of the problem, embodied
in the function $\epsilon(\omega)$, are separated from the geometric ones;
only these geometric aspects matter when studying deformations of $\Omega$.
On the other hand, standard theorems concerning the behavior of eigenvalues
of linear operators under perturbations can now be employed. After such a
plasmonic eigenvalue has been found for a particular geometry, {\em i.e.\/},
for some domain $\Omega$ with a given shape, that eigenvalue then determines 
the frequency $\omega_{\rm s}$ of a surface resonance through the equation 
$\epsilon = {\rm Re}\big(\epsilon(\omega_{\rm s})\big)$. For example, 
a half space with planar boundary $\partial\Omega$ yields $\epsilon = -1$, 
and the dielectric function can be taken as
\begin{equation}
	\epsilon(\omega) = 1 - \frac{\omega_{\rm p}^2}{\omega^2} 
\end{equation}  
for metallic materials with plasma frequency $\omega_{\rm p}$, describing 
the free electron motion. This then leads to the familiar expression 
$\omega_{\rm s} = \omega_{\rm p}/\sqrt{2}$ for the surface plasmon resonance
at a planar metal-vacuum interface in the quasistatic limit.~\cite{Raether88} 
Typical plasma frequencies for good conductors are on the order of 
$10^{16}$~s$^{-1}$.  

We now consider the response of a plasmonic eigenvalue $\epsilon$ to
a small deformation of $\Omega$. This deformation is modeled in terms 
of a shape function $a : \partial \Omega \to \R$ and a dimensionless 
parameter~$h$, such that the surface of the deformed volume is given by
\begin{equation}
	\partial\Omega(h) = \{\bfr + h a(\bfr) \, \bfn(\bfr) : \, 
	\bfr \in \partial\Omega\} \; .
\end{equation}	
We assume that one has solutions $\epsilon(h)$, $u(\bfr,h)$ to the problem
(\ref{EQ1}) -- (\ref{EQ3}) with $\Omega$ replaced by $\Omega(h)$, and that
$\epsilon(h)$ and $u(\bfr,h)$ depend differentiably on $h$. We denote the
derivatives in $h$ by a dot and write $\dot{\epsilon} = \dot{\epsilon}(0)$
and $\udot(\bfr) = \udot(\bfr,0)$ for brevity.  

In what follows we derive a system of equations for $\dot{\epsilon}$ and 
$\udot(\bfr)$, and therefrom an explicit expression for $\dot{\epsilon}$. To 
this end, we take the derivative of the system (\ref{EQ1}) -- (\ref{EQ3}) with 
respect to $h$. First, for any $\bfr\not\in\partial\Omega = \partial\Omega(0)$ 
one has $\bfr\not\in\partial\Omega(h)$ for $h$ sufficiently close to zero; 
therefore, Eq.~(\ref{EQ1}) readily yields 
\begin{equation}
	\Delta \udot = 0 \quad\text{in }\R^3\setminus\partial\Omega \; .
\label{EQ11}
\end{equation}
Next we differentiate Eq.~(\ref{EQ2}). In a more explicit form, this 
equation reads
\begin{equation}
	u_-(\bfr + h a(\bfr) \, \bfn(\bfr) , h) -
	u_+(\bfr + h a(\bfr) \, \bfn(\bfr) , h) = 0
\end{equation}
and hence gives
\begin{equation}
	a \bfn \cdot \bnab u_- + \udot_-
	- \left( a \bfn \cdot \bnab u_+ + \udot_+ \right) = 0
\end{equation}
or
\begin{eqnarray}
	\udot_- - \udot_+ & = & 
	a \bfn \cdot \bnab u_+ - a \bfn \cdot \bnab u_-
\nonumber \\	& = &
	(\epsilon - 1) a \partial_n u_- \; ,	 
\label{EQ22}
\end{eqnarray}
where Eq.~(\ref{EQ3}) has been used for eliminating $u_+$.

Similarly we differentiate Eq.~(\ref{EQ3}), keeping in mind that
$\partial_n u_\pm = \bfn \cdot \bnab u_\pm$, and taking into account that
$\bfn$ also depends on $h$. Thus, one has
\begin{equation}
 	(\bfn \cdot \bnab u_\pm)\dot{} = 
	\bfndot \cdot \bnab u_\pm + \bfn \cdot (\bnab u_\pm)\dot{} \; . 
\label{DER}
\end{equation}
The required derivatives $\bfndot$ and $\bfn \cdot (\bnab u_\pm)\dot{}$ 
are calculated in Appendix~\ref{App1}. According to Eq.~(\ref{RE1}), 
$\bfndot = -\nablad a$ coincides with the negative gradient of the shape 
function inside the surface $\partial\Omega$, while Eq.~(\ref{RE2}) 
expresses $\bfn \cdot (\bnab u_\pm)\dot{}$ in terms of the Laplace-Beltrami 
operator $\Deltad = \divd\nablad$ applied to $u_\pm$, and the mean curvature 
$H$ of $\partial\Omega$. Combining these results, we find 
\begin{eqnarray}
	& & 	
	(\bfn \cdot \bnab u_\pm)\dot{} 
\nonumber \\ 	& = &	
        -\nablad a \cdot \nablad u_\pm 
        - a \Deltad u_\pm + 2a H \partial_n u_\pm 
	+ \partial_n \udot_\pm 
\nonumber \\ 	& = & 
	- \divd (a\nablad u_\pm) 
	+ 2aH \partial_n u_\pm + \partial_n \udot_\pm \; . 	
\label{COM}
\end{eqnarray}
Thus, taking the $h$-derivative of Eq.~(\ref{EQ3}) leads to 
\begin{eqnarray}
	(\epsilon \partial_n u_- - \partial_n u_+)\dot{} & = &
	\dot{\epsilon} \partial_n u_- + \epsilon (\partial_n u_-)\dot{}
	- (\partial_n u_+)\dot{}
\nonumber \\	& = &
	\dot{\epsilon} \partial_n u_-
\nonumber \\ & & 
	- \divd\left(a (\epsilon \nablad u_- - \nablad u_+)\right) 
\nonumber \\ & & 
	+ 2 a H (\epsilon\partial_n u_- - \partial_n u_+) 
\nonumber \\ & & 
	+ \epsilon \partial_n \udot_- - \partial_n \udot_+ \; .			
\end{eqnarray}	
The l.h.s.\ now vanishes identically, and the third term on the r.h.s.\
vanishes because of Eq.~(\ref{EQ3}) itself. Moreover, since $u_+ = u_-$ 
on $\partial\Omega$ according to Eq.~(\ref{EQ2}), and $\nablad$ 
differentiates inside $\partial\Omega$ only, we can replace $\nablad u_+$
by $\nablad u_-$ in the second term, finally leaving us with 
\begin{equation}
	\epsilon \partial_n \udot_- - \partial_n \udot_+ =
	- \dot{\epsilon} \partial_n u_-
	+ (\epsilon - 1) \divd (a \nablad u_-) \; .
\label{IMP}
\end{equation}
In summary, the differentiation of the homogeneous system (\ref{EQ1})
-- (\ref{EQ3}) has led to the inhomogeneous system 
\begin{eqnarray}
	\Delta \udot & = & 0  \quad\text{in }\R^3\setminus\partial\Omega
\\
	\udot_- - \udot_+ & = & F \quad\text{on }\partial\Omega
\\
	\epsilon \partial_n \udot_- - \partial_n \udot_+ & = & 
	- \dot{\epsilon} \partial_n u_- + G
	\quad\text{on } \partial\Omega \; ,
\end{eqnarray}
where
\begin{eqnarray}
	F & = & (\epsilon - 1) a \partial_n u_-
\label{FOF}
\\
	G & = & (\epsilon - 1) \divd (a \nablad u_-) \; .
\label{FOG}
\end{eqnarray}
Since the homogeneous system has a nontrivial solution (the given $u$),
its inhomogeneous descendant can admit a solution only if the right hand 
sides $F$, $- \dot{\epsilon} \partial_n u_- + G$ satisfy a certain
compatibility condition. Since by assumption there is a solution (the
$h$-derivative of the given family $u(\bfr,h)$), this condition actually
is fulfilled, and yields an expression for $\dot{\epsilon}$. For deriving this
expression, we start from $\Delta u_- = \Delta \udot_- = 0$ in $\Omega$.
Green's formula then gives
\begin{eqnarray}
	0 & = & \int_\Omega \! \rd^3 r \, 
	\big( u_- \Delta \udot_- - (\Delta u_-) \udot_- \big)
\nonumber \\	& = &
	\int_{\partial\Omega} \! \rd^2 S \, 
	\big( u_- \partial_n \udot_- - (\partial_n u_-) \udot_- \big) \; ,
\label{GRF}
\end{eqnarray}
where $\rd^2 S$ denotes the surface area element. Thus, one has 
\begin{equation}
	\int_{\partial\Omega} \! \rd^2 S \, u_- \partial_n \udot_- =
	\int_{\partial\Omega} \! \rd^2 S \, (\partial_n u_-) \udot_- \; ; 
\end{equation}
a corresponding identity holds for $u_+$, $\udot_+$. Therefore,
\begin{eqnarray}
         & & 
	\int_{\partial\Omega} \! \rd^2 S \,
	\big( 
	u_- \epsilon \partial_n \udot_- - u_+ \partial_n \udot_+ 
	\big)
\nonumber \\	& = &
	\int_{\partial\Omega} \! \rd^2 S \,
	\big( 
	\epsilon (\partial_n u_-)\udot_- - (\partial_n u_+) \udot_+
	\big) \; .
\label{DIF}
\end{eqnarray}
Since $u_- = u_+$ on $\partial\Omega$, the l.h.s.\ becomes
\begin{equation}
	\int_{\partial\Omega} \! \rd^2 S \; u_- 
	\big( \epsilon \partial_n \udot_- - \partial_n \udot_+ \big)
	=
	\int_{\partial\Omega} \! \rd^2 S \; u_- 
	\big( -\dot{\epsilon} \partial_n u_- + G \big) \; .
\label{EX1}
\end{equation}
On the other hand, since $\partial_n u_+ = \epsilon \partial_n u_-$ on
$\partial\Omega$, the r.h.s.\ of Eq.~(\ref{DIF}) takes the form
\begin{equation}
	\int_{\partial\Omega} \! \rd^2 S \; \epsilon (\partial_n u_-)
	\left( \udot_- - \udot_+ \right) =
	\epsilon \int_{\partial\Omega} \! \rd^2 S \;
	(\partial_n u_-) F \; .	
\label{EX2}
\end{equation}
Equating these two expressions (\ref{EX1}) and (\ref{EX2}), and solving 
for $\dot{\epsilon}$, we obtain
\begin{equation}
	\dot{\epsilon} = \frac{\int_{\partial\Omega} \! \rd^2 S \; u_- G
	- \epsilon \int_{\partial\Omega} \! \rd^2 S \, (\partial_n u_-) F}
	{\int_{\partial\Omega} \! \rd^2 S \; u_- \partial_n u_-} \; . 
\end{equation}
Inserting the formulas (\ref{FOF}) and (\ref{FOG}) for $F$ and $G$,
and performing an integration by parts, the numerator reduces to 
\begin{eqnarray}
	& &
	\int_{\partial\Omega} \! \rd^2 S \;
	u_- (\epsilon - 1) \divd (a \nablad u_-)
\nonumber \\ 	
	& & 
	- \epsilon \int_{\partial\Omega} \! \rd^2 S \;
	(\partial_n u_-) (\epsilon - 1) a \partial_n u_-
\nonumber \\ & = &
	(1 - \epsilon) \int_{\partial\Omega} \! \rd^2 S \, a 
	\big( (\nablad u_-)^2 + \epsilon (\partial_n u_-)^2 \big) \; .			
\end{eqnarray}
Thus, the desired expression for the first-order change of the plasmonic
eigenvalue $\epsilon$ with the deformation strength~$h$ finally reads
\begin{equation}
	\dot{\epsilon} = (1-\epsilon) 
	\frac{\int_{\partial\Omega} \! \rd^2 S \; a
	\left( (\nablad u_-)^2 + \epsilon (\partial_n u_-)^2 \right)}
	{\int_{\partial\Omega} \! \rd^2 S \, u_- \partial_n u_-} \; .
\label{RES}
\end{equation}
This is the principal result of the present work. Observe that
$\Delta u_- = 0$ implies
\begin{equation}
	\int_{\partial\Omega} \! \rd^2 S \, u_- \partial_n u_-
	= \int_\Omega \! \rd^3 r \, (\nabla u_-)^2 \; ,	
\end{equation}
so that the denominator is positive. Of course, $\dot{\epsilon}$ can 
likewise be expressed entirely in terms of $u_+$:
\begin{equation}
	\dot{\epsilon} = (1-\epsilon) 
	\frac{\int_{\partial\Omega} \! \rd^2 S \; a
	\left( \epsilon (\nablad u_+)^2 + (\partial_n u_+)^2 \right)}
	{\int_{\partial\Omega} \! \rd^2 S \, u_+ \partial_n u_+} \; .
\label{VAR}
\end{equation}

\section{Applications}
\label{S_exam}

Consider an infinite half-space geometry with the dielectric medium filling
the volume $z < 0$, so that its boundary is given by the plane $z=0$. 
Let $\bk = (k_x, k_y)$ be a two-dimensional wave vector, with
$\kappa = \sqrt{k_x^2 + k_y^2}$. Solutions to Laplace's equation~(\ref{EQ1})
which vanish for $z \to \pm \infty$ then are given by
\begin{eqnarray}
  	u_+(\bfr) & = & \int \! \frac{\rd^2 \kappa}{(2 \pi)^2} \, A(\bk) \, 
	\re^{\ri \bk \cdot \mathbf{x} - \kappa z} 
\label{UPR}	\\
  	u_-(\bfr) & = & \int \! \frac{\rd^2 \kappa}{(2 \pi)^2} \, B(\bk) \, 
	\re^{\ri \bk \cdot \mathbf{x} + \kappa z} \; , 
\label{UMR}
\end{eqnarray}
with $\mathbf{x} = (x,y)$ and $\bfr = (x,y,z)$, and the continuity 
condition~(\ref{EQ2}) yields  $B(\bk) = A(\bk)$. Moreover, since $u_\pm$ 
is real, one has $A^*(\bk) = A(-\bk)$. At this point, the amplitudes
$A(\bk)$ describing the excitations at an exactly planar surface are
arbitrary. The normal and the in-plane derivatives at $z = 0$ then are
\begin{eqnarray}
  	\partial_n u_{\pm} & = & 
	\mp \int \! \frac{\rd^2 \kappa}{(2 \pi)^2} \, A(\bk) \, \kappa \,  
	\re^{\ri \bk \cdot \mathbf{x}} 
\\
  	\bnab_\partial u_{\pm} & = &
	\int \! \frac{\rd^2 \kappa}{(2 \pi)^2} \, A(\bk) \, (\ri \bk) \, 
	\re^{\ri \bk \cdot \mathbf{x}} \; ,
\end{eqnarray} 
respectively, so that the condition~(\ref{EQ3}) immediately provides the
well known eigenvalue $\epsilon = -1$ for this particular geometry. We
write
\begin{eqnarray}
        (\partial_n u_\pm)^2 & = &
	\int \! \frac{\rd^2 \kappa}{(2 \pi)^2} 
        \int \! \frac{\rd^2 \kappa'}{(2 \pi)^2}
        A(\bk) A^*(\bk') \, (\kappa \kappa') \, 
	\re^{\ri (\bk - \bk') \cdot \mathbf{x}} 
\nonumber \\
  	(\bnab_\partial u_\pm)^2 & = &
	\int \! \frac{\rd^2 \kappa}{(2 \pi)^2} 
        \int \! \frac{\rd^2 \kappa'}{(2 \pi)^2}
        A(\bk) A^*(\bk') \, (\bk \cdot \bk') \, 
	\re^{\ri (\bk - \bk') \cdot \mathbf{x}}
\nonumber
\end{eqnarray}
together with 
\begin{equation}
  	u_\pm \partial_n u_\pm = 
	\mp \int \! \frac{\rd^2 \kappa}{(2 \pi)^2} 
        \int \! \frac{\rd^2 \kappa'}{(2 \pi)^2}
        A(\bk) A^*(\bk') \, \kappa' \, 
	\re^{\ri (\bk - \bk') \cdot \mathbf{x}} \; ,
\nonumber
\end{equation}
giving
\begin{equation}
  	\int_{\partial\Omega} \! \rd^2 S \, u_\pm \partial_n u_\pm
        = \mp \int \! \frac{\rd^2 \kappa}{(2 \pi)^2} \, 
	|A(\bk)|^2 \kappa \; .
\end{equation}
Now we introduce a small deformation of the planar surface, described by some 
suitable shape function $a(\mathbf{x})$. The required amplitudes $A(\bk)$ 
consequently are determined by that deformation; the assumption that the 
exact potential $u_\pm(\bfr)$ can still be written in the form (\ref{UPR}) 
or (\ref{UMR}) constitutes the Rayleigh hypothesis.~\cite{FariasMaradudin83}   
Upon inserting the Fourier transform of the shape function, {\em i.e.\/}, 
\begin{equation}
  	a(\mathbf{x}) = \int \! \frac{\rd^2 \kappa}{(2 \pi)^2} \, 
	\widehat{a}(\bk) \, \re^{\ri \bk \cdot \mathbf{x}} \; ,
\end{equation}
und using $\epsilon = -1$ for the unperturbed eigenvalue, either 
Eq.~(\ref{RES}) or its variant (\ref{VAR}) readily yields
\begin{equation}
  	\dot{\epsilon} = 2 \, \frac{\int \! \frac{\rd^2 \kappa}{(2 \pi)^2} 
        \int \! \frac{\rd^2 \kappa'}{(2 \pi)^2} \, 
	\widehat{a}(\bk' - \bk) 
	A(\bk) A^*(\bk') \kappa \kappa' (\hat{\bk} \cdot \hat{\bk}' - 1)}
        {\int \! \frac{\rd^2 \kappa}{(2 \pi)^2} \, |A(\bk)|^2 \kappa} \; ,
\label{PEP}
\end{equation}
where $\hat{\bk}$ denotes the unit vector in the direction of $\bk$. 
In particular, if $a(\mathbf{x}) = c$, one has 
$\widehat{a}(\bk) = (2\pi)^2 c \delta(\bk)$, so that $\dot{\epsilon} = 0$: 
The plasmonic eigenvalue does not change when the entire surface plane is 
displaced by some amount~$c$. 

On the other hand, there exists an exact integral equation for determining 
the amplitudes $A(\bk)$ associated with a given surface deformation, derived 
by Farias and Maradudin on the basis of the Rayleigh 
hypothesis:~\cite{FariasMaradudin83}   
\begin{equation}
  	\frac{\epsilon + 1}{\epsilon - 1} A(\bk') 
        = \int \! \frac{\rd^2 \kappa}{(2 \pi)^2} \, 
        J(\kappa' - \kappa| \bk' - \bk) (1 - \hat{\bk} \cdot \hat{\bk}') 
        \kappa A(\bk)
\label{MAR}
\end{equation}
with
\begin{equation}
  	J(\alpha|\mathbf{q}) = \frac{1}{\alpha} \int \! \rd^2 S \, 
	\re^{- \ri \mathbf{q}\cdot\mathbf{x}} 
        \big( \exp(\alpha \, h a(\mathbf{x})) - 1 \big) \; ;
\label{TLE}
\end{equation}
this has been applied by Maradudin and Visscher to the study of particular 
perturbations of planar surfaces.~\cite{MaradudinVisscher85} Expanding the
latter expression~(\ref{TLE}) to first order in $h$ gives
\begin{equation}
  	J(\alpha|\mathbf{q}) = 
	h \widehat{a}(\mathbf{q}) + \mathcal{O}(h^2) \; ,
\end{equation}
so that Eq.~(\ref{MAR}) becomes
\begin{equation}
  	\frac{\epsilon + 1}{\epsilon - 1} A(\bk') 
        = h \int \! \frac{\rd^2 \kappa}{(2 \pi)^2} \, 
        \widehat{a}(\bk' - \bk) (1 - \hat{\bk} \cdot \hat{\bk}') 
        \kappa A(\bk)
\end{equation}
for sufficiently weak perturbations. Recalling that the unperturbed eigenvalue 
is $-1$, we insert $\epsilon = -1 + \mathcal{O}(h)$ into the denominator on 
the l.h.s. Multiplying both sides by $A^*(\bk')\kappa'$, integrating, and 
rearranging then yields  
\begin{equation}
	\epsilon = -1 + h \dot{\epsilon} + \mathcal{O}(h^2) \; ,
\end{equation}
with $\dot{\epsilon}$ indeed formally equal to the previous Eq.~(\ref{PEP}). 
Thus, our perturbative result is consistent with the formula~(\ref{MAR}).

A second example which allows one to confirm the validity of the perturbative
approach by analytical means is provided by the deformation of a dielectric 
sphere of radius~$R$ into a spheroid. In this case the unperturbed potential 
is written as
\begin{eqnarray}
	u_+(\bfr) & = & \sum_{\ell = 1}^\infty 
	A_\ell r^{-(\ell+1)} Y_0^\ell (\vartheta) 
\nonumber \\
	u_-(\bfr) & = & \sum_{\ell = 0}^\infty 
	B_\ell r^\ell Y_0^\ell (\vartheta) \; ,
\end{eqnarray}
valid for $r \ge R$ and $r \le R$, respectively. Here $Y_m^\ell$ denote the 
familiar spherical harmonics. The restriction to these basis functions with 
$m = 0$, which do not depend on the azimuthal angle $\varphi$, confines us to
deformations which preserve the rotation symmetry around the $z$-axis.  
From condition~(\ref{EQ2}) one gets $B_\ell = A_\ell R^{-(2\ell + 1)}$, 
and Eq.~(\ref{EQ3}) then leads to the eigenvalues
\begin{equation}
  	\epsilon_\ell = - \frac{\ell + 1}{\ell} \; .
\end{equation}
In particular, $\epsilon_1 = -2$ characterizes the dipole 
resonance.~\cite{BohrenHuffman98}

We deform the sphere into a spheroid oriented along the $z$-axis, employing 
the shape function
\begin{equation}
  	a(\vartheta,\varphi) = R\cos^2(\vartheta) \; .
\label{SHF}
\end{equation}
Here the choice of the sphere's radius $R$ as the scale of the deformation
is a matter of convenience. The three degenerate dipole modes of the sphere
are shifted when the deformation strength~$h$ adopts nonzero values, such 
that the mode associated with the $z$-axis splits off from the two others. 
For estimating the corresponding change of $\epsilon_1$ for small $h$, 
we use
\begin{equation}
	u_{1-}(\bfr) = r Y_0^1(\vartheta) 
	= r \sqrt{\frac{3}{4\pi}} \cos \vartheta \; ,
\label{UDM}
\end{equation}
and calculate
\begin{eqnarray*}
	(\partial_n u_{1-})^2 & = & \frac{3}{4\pi} \cos^2 \vartheta	
\\	
	(\nablad u_{1-})^2 & = & \frac{3}{4\pi} \sin^2 \vartheta
\\
	u_{1-}\partial_n u_{1-} & = & \frac{3}{4\pi} R \cos^2 \vartheta
\end{eqnarray*}
on $\partial\Omega$, giving
\begin{eqnarray}
	\int_{\partial\Omega} \! \rd^2 S \, a (\partial_n u_{1-})^2 
	& = & \frac{3}{5} R^3
\nonumber \\
	\int_{\partial\Omega} \! \rd^2 S \, a (\nablad u_{1-})^2 
	& = & \frac{2}{5} R^3
\nonumber \\ 
	\int_{\partial\Omega} \! \rd^2 S \, u_{1-}\partial_n u_{1-} 
	& = & R^3 \; .
\end{eqnarray}
Plugged into Eq.~(\ref{RES}), this yields
\begin{equation}
	\dot{\epsilon}_1 = -\frac{12}{5} \; .
\label{PRE}
\end{equation}
This easily obtainable result can be checked against the exactly known
expression for the dipolar $z$-mode of a spheroid:~\cite{BohrenHuffman98}
\begin{equation}
	\epsilon_1 = 1 - \frac{1}{L} \; ,
\label{EP1}
\end{equation}	
where $L$ denotes the so-called depolarization coefficient,
given by~\cite{LandauLifshitzVIII}
\begin{equation}
	L(e) = \begin{cases}
       	\frac{1-e^2}{e^3}\big( {\rm artanh}(e) - e \big) \; ,
       	& h > 0
 \\
        \frac{1+e^2}{e^3}\big( e - \arctan(e) \big) \; ,
        & h < 0 
        \end{cases}
\end{equation}
with
\begin{equation}
  	e^2 = \begin{cases}
       	1 - \frac{1}{(1+h)^2} =  2h + \mathcal{O}(h^2) \; , 
       	& h > 0
\\
       	\frac{1}{(1+h)^2} - 1 = -2h + \mathcal{O}(h^2) \; , 
       	& h < 0 \; .
       	\end{cases}
\label{EXE}
\end{equation}
Thus, $e^2$ becomes small with $h$, allowing us to expand $L(e)$ as
\begin{equation}
  	L(e) = \begin{cases}
        \frac{1}{3} - \frac{2}{15} e^2 + \mathcal{O}(e^4) \; , 
        & h > 0
\\      \frac{1}{3} + \frac{2}{15} e^2 + \mathcal{O}(e^4) \; , 
        & h < 0 \; .
       \end{cases}
\end{equation}
Exploiting Eq.~(\ref{EXE}) one expresses $L$ in terms of $h$, finding  
\begin{equation}
	L(h) = \frac{1}{3} - \frac{4}{15}h + \mathcal{O}(h^2) \; .
\label{EXL}
\end{equation}
This tells us that $\epsilon_1$, as given by Eq.~(\ref{EP1}), should be
expanded in powers of $(L - 1/3)$:
\begin{equation}
  	\epsilon_1 = -2 + 9\left(L-\frac{1}{3}\right)
  	+ \mathcal{O}\left(\left(L-\frac{1}{3}\right)^2\right) \; .
\end{equation}
Inserting Eq.~(\ref{EXL}), we obtain
\begin{equation}
	\epsilon_1 = -2 - \frac{12}{5}h + \mathcal{O}(h^2) \; ,
\end{equation}
in accordance with the result~(\ref{PRE}) of perturbation theory.

\begin{figure}
\includegraphics[scale=0.5]{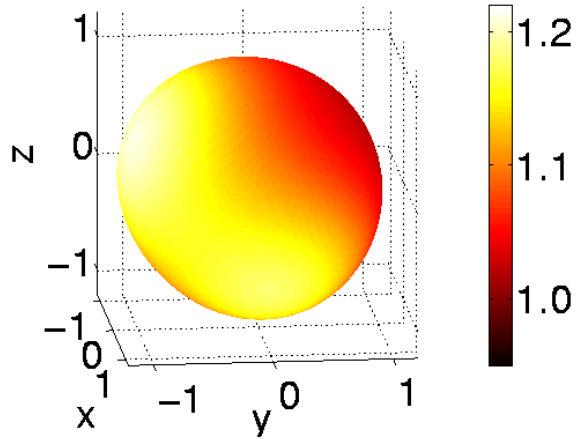}
\caption{(Color online) Sketch of the particular perturbed 
	nanosphere~(\ref{NSP}) employed as example for demonstrating the 
	accuracy of the perturbative approach. The deviation of the scaled
	surface radii $r(\vartheta,\varphi) / R$ from their ideal value~$1$ 
	is visualized here for $h = 0.2$.} 
\label{F_1}
\end{figure}

The following third example concerns a deformed nanosphere which does not 
admit an analytical solution in closed form, so that the accuracy achieved 
by first-order perturbation theory has to be ascertained by comparison with 
numerical calculations. This is exactly the type of application we have in 
mind, since here Eqs.~(\ref{RES}) and (\ref{VAR}) provide a quick and 
reliable estimate of the possibly detrimental consequences of geometrical 
imperfections,~\cite{deWaeleEtAl07} without hard requirements on 
computational resources. We assume that the ideal sphere is distorted by 
two Gaussian protrusions, and parametrize its surface as 
\begin{equation}
	a(\vartheta,\varphi) = R \sum_{i=1,2} f_i 
	\exp\left(
	-\frac{{ \rm dist}(\vartheta_i,\varphi_i;\vartheta,\varphi)^2}
	{2 w_i^2} 
	\right) \; ,
\label{NSP}	
\end{equation}
with ${\rm dist}(\vartheta_i,\varphi_i;\vartheta,\varphi)$ denoting the
dimensionless Euclidean distance between the two points on the unit sphere 
specified by the angles $\vartheta_i,\varphi_i$ and $\vartheta,\varphi$, 
respectively. The protrusions' parameters are chosen, somewhat arbitrarily, 
as $f_1 = 1.0$, $f_2 = 0.8$ for their amplitudes, $w_1 = 0.7$, $w_2 = 0.65$
for their Gaussian widths, and $\vartheta_1 = 1.1$, $\vartheta_2 = 2.2$
and $\varphi_1 = 4.9$, $\varphi_2 = 0.1$ for their locations. 
Figure~\ref{F_1} depicts the resulting deformed sphere for the value
$h = 0.2$ of the overall perturbation strength.
   
\begin{figure}
\includegraphics[scale=0.6]{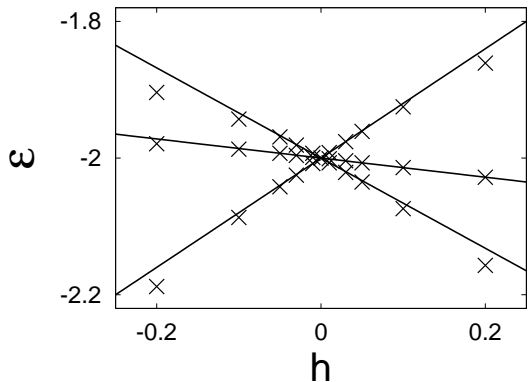}
\caption{Numerically computed plasmonic eigenvalues for the dipole modes 
	of the deformed nanosphere depicted in Fig.~\ref{F_1} (crosses), 
	in comparison with the results of first-order perturbation theory 
	(full lines).} 
\label{F_2}
\end{figure}

This example also illustrates a further important issue. While the dipole
modes of a perfect sphere are threefold degenerate, this degeneracy is removed 
entirely by the distortion~(\ref{NSP}). Thus, there now are three branches of
dipole-like eigenvalues, and our previous analysis applies to each branch 
separately, provided the starting point is chosen appropriately: The 
expressions~(\ref{RES}) and (\ref{VAR}) refer to the individual branches, if 
the proper linear combinations of the unperturbed degenerate modes are 
inserted. The problem how to find these linear combinations is solved in 
Appendix~\ref{App2}. Basically, the numerator of the formula~(\ref{RES})
defines a quadratic form of the eigenmodes (see Eq.~(\ref{QUA})); the 
required proper linear combinations of the unperturbed modes are those which 
diagonalize this form. Their determination, and the evaluation of the ensuing 
surface integrals determining $\dot{\epsilon}$ for all three branches, does
not demand much numerical effort. Figure~\ref{F_2} shows the perturbative 
results for deformation strengths $-0.25 < h < + 0.25$ in comparison with 
data obtained by nonperturbative numerical computations.~\cite{RuetingUecker09}
Quite remarkably, first-order perturbation theory still produces excellent 
results when the shape variation already is quite substantial, that is, for 
values of $h$ up to $0.1$; even for $h$ as large as $0.2$ one obtains good 
estimates.    
 
\section{Concluding remarks}
\label{S_conc}

The first-order perturbative expression~(\ref{RES}) or (\ref{VAR}) for the 
shift of a plasmonic eigenvalue clearly has a limited range of applicability,
insofar as it is restricted to sufficiently weak deformations, but it stands
out because of its generality, and easy use. Together with similar results
for higher derivatives,~\cite{Grieser09} one obtains a formal perturbation
series
\begin{equation}
	\epsilon(h) \sim \epsilon + h \dot{\epsilon} 
	+ \frac{1}{2}h^2 \ddot{\epsilon} + \ldots \; . 
\end{equation}
Above we have simply assumed the existence of $\epsilon(h)$ and $u(\bfr,h)$.
Given a solution at $h = 0$, this existence actually follows from the
reformulation of the problem in terms of Dirichlet-to-Neumann 
operators,~\cite{GrieserRueting09} and from standard perturbation theory 
for eigenvalues. If $\epsilon$ is $p$-fold degenerate (as in the case of 
the sphere), then one has $p$ branches $\epsilon^{(i)}$, $u^{(i)}$; 
$i = 1, \ldots , p$. The analysis of Sec.~\ref{S_pert} then applies to each 
branch separately, with $\dot{\epsilon}^{(i)}$ being given by Eq.~(\ref{RES}) 
or (\ref{VAR}) with $u = u^{(i)}$. Here $u^{(1)}, \ldots , u^{(p)}$ form 
a basis of the subspace of degenerate solutions at $h = 0$, determined such 
that they diagonalize the quadratic form given by Eq.~(\ref{QUA}).

The three examples we have given in Sec.~\ref{S_exam} vary in character, 
the first two recovering known analytical results and thus confirming the 
correctness of our formal line of reasoning. The third example, summarized 
by Figs.~\ref{F_1} and \ref{F_2}, demonstrates the utility of our approach
for practical purposes. Here we have dealt with an asymmetrically deformed 
nanosphere, and obtained fairly good estimates for the shifted dipole modes.  
If we assume that tolerances on the order of 5\% can be met in nanosphere 
fabrication, first-order perturbation theory thus allows one to quantify
the effects of a large variety of possible shape fluctuations with both 
sufficent accuracy and only small numerical effort.

\begin{acknowledgments}
This work was supported in part by the Deutsche Forschungsgemeinschaft
under grant No.\ KI 438/8. S.-A.~B.\ gratefully acknowledges a fellowship 
from the Deutsche Akademie der Naturforscher Leopoldina. 
\end{acknowledgments}

\begin{appendix}

\section{Auxiliary calculations}
\label{App1}

In this appendix we compute the derivatives entering into the evaluation 
of Eq.~(\ref{DER}) en route to the important result~(\ref{IMP}). This 
calculation invokes some notions from differential geometry.~\cite{doCarmo76} 
We fix a  point $\bfr_0 \in \partial\Omega = \partial\Omega(0)$ and 
parametrize $\partial\Omega$ near $\bfr_0$ as $\bfr = \bfr(v,w)$, with 
$\bfr_0 = \bfr(0,0)$. It is assumed that the parametrization is such that 
the tangent vectors $\bfr_v = \partial_v \bfr$ and $\bfr_w = \partial_w \bfr$
are orthogonal, and have unit length at $\bfr_0$, and such that the $v$-line
({\em i.e.\/}, $v \mapsto \bfr(v,0)$) and the $w$-line through $\bfr_0$ have
normal vectors parallel to $\bfn$ at $\bfr_0$ ({\em i.e.\/}, have 
vanishing geodesic curvature). 

The curvatures of the $v$-line and the $w$-line are 
$\kappa_v = \bfr_{vv} \cdot \bfn$ and $\kappa_w = \bfr_{ww} \cdot \bfn$,
respectively. Considering $\bfn$ as a function of $v$, $w$, one has  
\begin{equation}
	\bfr_v \cdot \bfn = 0
\end{equation}
and hence
\begin{equation}
	\bfr_{vv} \cdot \bfn + \bfr_v \cdot \bfn_v = 0 \; , 
\end{equation}
giving
\begin{equation}
	\kappa_v = -\bfr_v \cdot \bfn_v \; ;
\label{KAV}
\end{equation}
analogously, $\kappa_w = -\bfr_w \cdot \bfn_w$.

Note that orthonormality of $\bfr_v$ and
$\bfr_w$ holds only at $\bfr_0$, not at nearby points. In the following 
calculations we always evaluate at $h = 0$ after differentiating in $h$, 
and at $\bfr$ = $\bfr_0$ after differentiating in $v$, $w$.   

We now derive an expression for $\bfndot$. The surface $\partial\Omega(h)$ 
is parametrized by
\begin{equation}
	\bfr(v,w,h) = \bfr(v,w) + h a(v,w) \bfn(v,w) \; ,
\end{equation}
where we abuse notation by considering $a$ and $\bfn$ as functions of 
$v$, $w$. Its unit normal is $\bfn = \bfN / \|\bfN\|$, where 
$\bfN =\bfr_v \times \bfr_w$. At $\bfr_0$ one has $\bfn = \bfN$ by
assumption. Differentiating, one gets
\begin{eqnarray}
	\dot{\bfn} & = & \frac{\dot{\bfN}}{\| \bfN \|} -
	\frac{\bfN}{\| \bfN \|^2}  
	\left( \frac{\bfN}{\| \bfN \|} \cdot \dot{\bfN} \right)
\nonumber \\	& = &
	\frac{1}{\| \bfN \|} 
	\left( \dot{\bfN} - \bfn (\bfn \cdot \dot{\bfN} ) \right) \; ,
\label{TAN}
\end{eqnarray}
which states that $\dot{\bfn}$ is determined by the tangential part of
$\dot{\bfN}$. Moreover,
\begin{equation}
	\dot{\bfN} = \dot{\bfr}_v \times \bfr_w + \bfr_v \times \dot{\bfr}_w 
	\; .
\label{NDT}	
\end{equation}
Since $\dot{\bfr} = a\bfn$, one has
\begin{equation}
	\dot{\bfr}_v = a_v \bfn + a \bfn_v \; .
\end{equation}	 
At $\bfr_0$, the identity $\bfn = \bfr_v \times \bfr_w$ implies
$\bfn \times \bfr_w = -\bfr_v$ and $\bfr_v \times \bfn = -\bfr_w$.
In addition, 
$\bfn_v \times \bfr_w = (\bfn_v \cdot \bfr_v) \bfr_v \times \bfr_w$, because 
the component of $\bfn_v$ parallel to $\bfr_w$ drops out of the vectorial 
product. (Observe that $\bfn \cdot \bfn = 1$ implies $\bfn_v \cdot \bfn = 0$,
so that $\bfn_v$ lies in the span of $\bfr_v$ and $\bfr_w$.) Putting this 
together and using Eq.~(\ref{KAV}), one finds
\begin{eqnarray}
	\dot{\bfr}_v \times \bfr_w & = &
	\left( a_v \bfn + a \bfn_v \right) \times \bfr_w
\nonumber \\	& = &
	- a_v \bfr_v + a(\bfn_v \cdot \bfr_v) \bfn
\nonumber \\	& = &	
	- a_v \bfr_v - a \kappa_v \bfn \; ;
\label{CR1}
\end{eqnarray}
analogously,
\begin{equation}
	\bfr_v \times \dot{\bfr}_w =
	- a_w \bfr_w - a \kappa_w \bfn \; .	
\label{CR2}
\end{equation}
Inserting Eqs.~(\ref{CR1}) and (\ref{CR2}) into Eq.~(\ref{NDT}), we 
arrive at
\begin{equation}
	\dot{\bfN} = -\left( a_v \bfr_v + a_w \bfr_w \right)
	- a \left( \kappa_v + \kappa_w \right) \bfn \; .
\end{equation}
Now we observe that
\begin{equation}
	a_v \bfr_v + a_w \bfr_w = \nablad a
\end{equation}
is the gradient of $a$ as a function in the surface $\partial\Omega$. 	
Since according to Eq.~(\ref{TAN}) the desired derivative $\dot{\bfn}$
is the tangential component of $\dot{\bfN}$, we have the compact result 
\begin{equation}
	\dot{\bfn} = -\nablad a \; .
\label{RE1}
\end{equation}
Next we calculate $\bfn \cdot (\bnab u)\dot{}$ for $u=u_+$ or $u=u_-$
at $\bfr_0$. Starting from
\begin{eqnarray}
	(\bnab u)\dot{}\,(\bfr) & = & 
	\left. \frac\partial{\partial h} \right|_{h=0} 
	(\bnab u)(\bfr + h a(\bfr)\bfn(\bfr),h) 
\nonumber \\	& = &
	a (\bfn\cdot\bnab) \bnab u (\bfr) 
	+ \bnab \udot(\bfr)
\end{eqnarray}
one finds
\begin{equation}
	\bfn \cdot (\bnab u)\dot{} = 
	a(\bfn \cdot \bnab)^2 u + \partial_n \udot \; . 
\end{equation}
To simplify notation we assume that coordinates in $\R^3$ are chosen such
that $\bfn = (0,0,1)$ at $\bfr_0$, reducing the first term on the r.h.s.\
to $a \partial_{x_3}^2 u$ at $\bfr_0$. Because $\Delta u = 0$, we then have,
at $\bfr_0$,
\begin{equation}
	\bfn \cdot (\bnab u)\dot{} = 
	-a \left(\partial_{x_1}^2   + \partial_{x_2}^2\right) u  
	+ \partial_n \udot \; .
\label{NGU}
\end{equation}
This has to be expressed in terms of derivatives of $u$ inside the surface.
We choose the surface coordinates as $v = x_1$, $w = x_2$, but still write
$\bfr = \bfr(v,w)$. Differentiating $u$ twice with respect to $v$,
\begin{eqnarray}
	\partial_v u(\bfr(v,w)) & = &
	(\bfr_v \cdot \bnab) u (\bfr(v,w)) 
\nonumber \\	
	\partial_v^2 u(\bfr(v,w)) & = &
	(\bfr_v \cdot \bnab)^2 u (\bfr(v,w))
\nonumber \\ & &
	+ (\bfr_{vv} \cdot \bnab) u (\bfr(v,w)) \; ,
\end{eqnarray}
and observing $\bfr_v \cdot \bnab = \partial_{x_1}$ and
$\bfr_{vv} = \kappa_v \bfn$, we get
\begin{equation} 
	\partial_v^2 u = \partial_{x_1}^2 u + \kappa_v \partial_n u \; .
\end{equation}	
Combining this with the analogous equation for $x_2$ gives
\begin{equation}
	\left(\partial_{x_1}^2 + \partial_{x_2}^2\right) u
	= \left( \partial_v^2 + \partial_w^2 \right) u
	- \left( \kappa_v + \kappa_w \right) \partial_n u \; . 
\end{equation}
Introducing the mean curvature $H$ of $\partial\Omega$,
\begin{equation}
	H = \frac{1}{2}\left( \kappa_v + \kappa_w \right) \; ,
\end{equation}
and identifying 
\begin{equation}
	\Deltad u = \left( \partial_v^2 + \partial_w^2 \right) u	
\end{equation}
as the Laplace-Beltrami operator of the surface $\partial\Omega$,
Eq.~(\ref{NGU}) takes the final form
\begin{equation}
	\bfn \cdot (\bnab u)\dot{} = 
	-a \Deltad u + 2a H \partial_n u + \partial_n \udot \; .
\label{RE2}
\end{equation}
While we had used a special coordinate system for simplifying the derivation, 
the Laplace-Beltrami operator is defined independent of the choice of 
coordinates as $\Deltad = \divd\nablad$. Here $\divd$ is the divergence 
inside the surface, given by the negative adjoint of the gradient with 
respect to the surface volume element. This is exploited in Eq.~(\ref{COM}).

The fact that the two results (\ref{RE1}) and (\ref{RE2}) obtained in this
appendix are expressed invariantly, that is, without reference to coordinates,
is essential for their use in Sec.~\ref{S_pert}.

\section{Perturbation theory for degenerate modes}
\label{App2}

When dealing with the splitting of degenerate eigenvalues in response to 
some deformation, as exemplified in Fig.~\ref{F_2}, the question emerges 
which linear combinations of the unperturbed modes must be inserted into 
Eqs.~(\ref{RES}) and (\ref{VAR}) in order to obtain the different branches. 
For finding these proper linear combinations, we start with the following 
assertion: If $u_\pm^{(1)}$, $u_\pm^{(2)}$ are two nondegenerate solutions 
to the system~(\ref{EQ1}) -- (\ref{EQ3}), with eigenvalues 
$\epsilon^{(1)} \ne \epsilon^{(2)}$, then
\begin{equation}
	\int_{\partial\Omega} \! \rd^2 S \; u_-^{(1)} \partial_n u_-^{(2)} =
	\int_{\partial\Omega} \! \rd^2 S \; u_-^{(2)} \partial_n u_-^{(1)}
	= 0 \; .
\label{ASE}
\end{equation}
This is shown by first exploiting Eq.~(\ref{EQ3}), and writing
\begin{equation}
	\epsilon^{(2)} \partial_n u_-^{(2)} - \partial_n u_+^{(2)} = 0 
	\quad\text{on } \partial\Omega \; .
\end{equation}
Multiplying by $u_-^{(1)}$ and integrating, one arrives at
\begin{eqnarray}
	\epsilon^{(2)} \int_{\partial\Omega} \! \rd^2 S \;
	u_-^{(1)} \partial_n u_-^{(2)}
	& = &
	\int_{\partial\Omega} \! \rd^2 S \;
	u_-^{(1)} \partial_n u_+^{(2)}
\nonumber\\ 	& = &
	\int_{\partial\Omega} \! \rd^2 S \;
	u_+^{(1)} \partial_n u_+^{(2)} \; ,	
\label{SU1}
\end{eqnarray}
where Eq.~(\ref{EQ2}) has been used. In the same manner one also finds
\begin{equation}
	\epsilon^{(1)} \int_{\partial\Omega} \! \rd^2 S \;
	u_-^{(2)} \partial_n u_-^{(1)}
	=
	\int_{\partial\Omega} \! \rd^2 S \;
	u_+^{(2)} \partial_n u_+^{(1)} \; .	
\label{SU2}
\end{equation}
Moreover, one has
\begin{eqnarray}
	\int_{\partial\Omega} \! \rd^2 S \;
	u_-^{(1)} \partial_n u_-^{(2)}
	& = &
	\int_\Omega \! \rd^3 r \; \nabla \cdot 
	\big( u_-^{(1)} \nabla u_-^{(2)} \big)
\nonumber\\	& = &
	\int_\Omega \! \rd^3 r \; \nabla u_-^{(1)} \cdot \nabla u_-^{(2)}
\end{eqnarray}
by virtue of Eq.~(\ref{EQ1}). Hence, we deduce
\begin{equation}
	\int_{\partial\Omega} \! \rd^2 S \;
	u_-^{(1)} \partial_n u_-^{(2)}
	=
	\int_{\partial\Omega} \! \rd^2 S \;
	u_-^{(2)} \partial_n u_-^{(1)} \; ;
\label{SYM}
\end{equation}
similarly,
\begin{equation}
	\int_{\partial\Omega} \! \rd^2 S \;
	u_+^{(1)} \partial_n u_+^{(2)}
	=
	\int_{\partial\Omega} \! \rd^2 S \;
	u_+^{(2)} \partial_n u_+^{(1)} \; .
\end{equation}
Therefore, subtracting Eq.~(\ref{SU2}) from Eq.~(\ref{SU1}) readily yields 	
\begin{equation}
	\big( \epsilon^{(2)} - \epsilon^{(1)} \big)
	\int_{\partial\Omega} \! \rd^2 S \; u_-^{(1)} \partial_n u_-^{(2)}
	= 0 \; .
\end{equation}	
Since $\epsilon^{(2)} \ne \epsilon^{(1)}$ by assumption, this, together with 
Eq.~(\ref{SYM}), was to be demonstrated.

Next, we consider two such branches of solutions which depend on a 
deformation strength~$h$ and are degenerate only for vanishing deformation,
$\epsilon^{(1)} = \epsilon^{(2)} = \epsilon$ for $h = 0$, while
$\dot{\epsilon}^{(1)} \ne \dot{\epsilon}^{(2)}$. Since then the 
assertion~(\ref{ASE}) holds for any $h \ne 0$, continuity demands that it 
is also valid for $h = 0$: It is this requirement~(\ref{ASE}) which singles 
out the proper linear combinations of the degenerate unperturbed modes,
when re-tracing the split modes back to the point of degeneracy. Focusing 
now on $h = 0$, and observing
\begin{equation}
	0 = \int_\Omega \! \rd^3 r \, 
	\big( u_-^{(1)} \Delta \udot_-^{(2)} 
	    - (\Delta u_-^{(1)}) \udot_-^{(2)} \big) \; ,
\end{equation}
virtually the same steps that also lead from Eq.~(\ref{GRF}) to Eq.~(\ref{RES})
then result in
\begin{eqnarray}
	& & 
	\dot{\epsilon}^{(2)} \int_{\partial\Omega} \! \rd^2 S \;
	u_-^{(1)} \partial_n u_-^{(2)}
\nonumber \\	& = & 
	\int_{\partial\Omega} \! \rd^2 S \; 
	\big( u_-^{(1)} G^{(2)} 
	- \epsilon (\partial_n u_-^{(1)}) F^{(2)} \big)
\nonumber \\	& = & 
	(1 - \epsilon) \, M(u^{(1)}, u^{(2)}) \; , 	 		
\label{CON}
\end{eqnarray}
where we have introduced the expression
\begin{equation}
	M(u^{(1)}, u^{(2)}) = \int_{\partial\Omega} \! \rd^2 S \; a
	\big( \nablad u_-^{(1)} \cdot \nablad u_-^{(2)}
	+ \epsilon \partial_n u_-^{(1)} \partial_n u_-^{(2)} \big) \; .	
\label{QUA}
\end{equation}
Because the l.h.s.\ of Eq.~(\ref{CON}) vanishes, so does the r.h.s. Excluding 
the particular value $\epsilon = 1$, we deduce $M(u^{(1)}, u^{(2)}) = 0$. This 
requirement finally dictates how to proceed in the general case: Let us assume 
that an eigenvalue $\epsilon$ to the system~(\ref{EQ1}) -- (\ref{EQ3}) is 
$p$-fold degenerate,  with eigenmodes $v^{(k)}$, where $k = 1, \ldots, p$. 
Then $M(v^{(k)}, v^{(\ell)})$ is a symmetric $p \times p$ -matrix and hence 
possesses $p$ orthonormal eigenvectors $\eta^{(j)}_\ell$ (where the lower
index $\ell$ refers to the components) with eigenvalues $d^{(j)}$, such that
\begin{equation}
	\sum_{\ell=1}^p M(v^{(k)}, v^{(\ell)}) \, \eta^{(j)}_\ell 
	= d^{(j)} \eta^{(j)}_k
\label{EVE}
\end{equation}
and
\begin{equation}
	\sum_{k=1}^p \eta^{(i)}_k \eta^{(j)}_k = \delta_{i,j} \; ,
\end{equation}
employing the usual Kronecker delta $\delta_{i,j}$. Therefore, multiplying 
Eq.~(\ref{EVE}) by $\eta^{(i)}_k$ and summing over $k$, one gets
\begin{equation}
	\sum_{\ell,k=1}^p 
	M( \eta^{(i)}_k v^{(k)}, \eta^{(j)}_\ell v^{(\ell)}) 
	= d^{(j)} \delta_{i,j} \; .
\end{equation}
Thus, setting
\begin{equation}
	u^{(i)} = \sum_{k=1}^p \eta^{(i)}_k v^{(k)} \; ,
\label{LIN}
\end{equation}
we have $M(u^{(i)},u^{(j)}) = 0$ for $i \ne j$, as required. This 
Eq.~(\ref{LIN}) therefore specifies the desired proper linear combinations 
of the unperturbed, degenerate eigenmodes. 
 
\end{appendix}

\end{document}